\documentclass[aps,preprint,amsmath,amssymb]{revtex4}
\usepackage{graphicx}
\begin{document}

\title{Analysis of anomalous quartic $WWZ\gamma$ couplings in $\gamma p$ collision at the LHC}

\author{A. Senol}
\email[]{senol_a@ibu.edu.tr} \affiliation{Department of Physics,
Abant Izzet Baysal University, 14280, Bolu, Turkey}

\author{M. K\"{o}ksal}
\email[]{mkoksal@cumhuriyet.edu.tr} \affiliation{Department of
Physics, Cumhuriyet University, 58140, Sivas, Turkey}

\begin{abstract}
Gauge boson self-couplings are exactly determined by the non-Abelian
gauge nature of the Standard Model (SM), thus precision measurements
of these couplings at the LHC provide an important opportunity to
test the gauge structure of the SM and the spontaneous symmetry
breaking mechanism. It is a common way to examine the physics of
anomalous quartic gauge boson couplings via effective Lagrangian
method. In this work, we investigate the potential of the process
$pp\rightarrow p\gamma p\rightarrow p W Z q X$  to analyze anomalous
quartic $WWZ\gamma$ couplings by two different CP-violating and
CP-conserving effective Lagrangians at the LHC. We calculate $95\%$
confidence level limits on the anomalous coupling parameters with
various values of the integrated luminosity. Our numerical results
show that the best limits obtained on the anomalous couplings
$\frac{k_{0}^{W}}{\Lambda^{2}}$, $\frac{k_{c}^{W}}{\Lambda^{2}}$,
$\frac{k_{2}^{m}}{\Lambda^{2}}$ and $\frac{a_{n}}{\Lambda^{2}}$ at
$\sqrt{s}=14$ TeV and an integrated luminosity of $L_{int}=100$
fb$^{-1}$ are $[-1.37;\, 1.37]\times 10^{-6}$ GeV$^{-2}$, $[-1.88;
\, 1.88]\times 10^{-6}$ GeV$^{-2}$, $[-6.55; \, 6.55]\times 10^{-7}$
GeV$^{-2}$ and $[-2.21;\,2.21]\times 10^{-6}$ GeV$^{-2}$,
respectively. Thus, $\gamma p$ mode of photon-induced reactions at
the LHC highly improves the sensitivity limits of the anomalous
coupling parameters $\frac{k_{0}^{W}}{\Lambda^{2}}$,
$\frac{k_{c}^{W}}{\Lambda^{2}}$, $\frac{k_{2}^{m}}{\Lambda^{2}}$ and
$\frac{a_{n}}{\Lambda^{2}}$.

\end{abstract}

\maketitle

\section{Introduction}

The SM has been tested with many important experiments and it has
been demonstrated to be quite successful, particularly after the
discovery of a particle consistent with the Higgs boson with a mass
of about $125$ GeV \cite{higgs1,higgs2}. Nevertheless, some of the
most fundamental questions still remain unanswered. Especially, the
strong CP problem, neutrino oscillations and matter - antimatter
asymmetry have not been adequately clarified by the SM. It is
expected to find answers to these problems of new physics beyond the
SM. One of the ways of investigating new physics is to examine
anomalous gauge boson interactions determined by non-Abelian
$SU_{L}(2)\times U_{Y}(1)$ gauge symmetry. Therefore, research on
these couplings with a high precision can either confirm the gauge
symmetry of the SM or give some hint for new physics beyond the SM.
Any deviation of quartic couplings of the gauge bosons from the
expected values would imply the existence of new physics beyond the
SM. It is mostly common to examine new physics in a model
independent way via the effective Lagrangian method. This method is
expressed by high-dimensional operators which lead to anomalous
quartic gauge couplings. These high-dimensional operators do not
generate new trilinear vertices. Thus, genuine quartic gauge
couplings can be independently investigated from new trilinear gauge
couplings.

In the literature, the anomalous quartic gauge boson couplings are
commonly examined by two different CP-conserving and CP-violating
effective Lagrangians. The first one, CP-violating effective
Lagrangian is defined by \cite{lag1}

\begin{eqnarray}
\textit{L}_{n}=\frac{i \pi \alpha}{4\Lambda^{2}} a_{n} \epsilon_{ijk} W_{\mu \alpha}^{(i)}W_{\nu}^{(j)} W^{(k)\alpha} F^{\mu\nu}
\end{eqnarray}
where $F^{\mu \nu}$ is the electromagnetic field strength tensor,
$\alpha$ is the electroweak coupling constant, $a_{n}$ is the
strength of the parametrized anomalous quartic coupling and
$\Lambda$ stands for new physics scale. The anomalous $W W Z \gamma$ vertex function generated by above effective
Lagrangian is given in Appendix.

Second, the CP-conserving effective operators can be written by
using the formalism of Ref. \cite{lhc}. There are fourteen effective
photonic operators with respect to the anomalous quartic gauge
couplings, and they are defined by 14 independent couplings
$k_{0,c}^{w,b,m},k_{1,2,3}^{w,m}$ and $k_{1,2}^{b}$ which are all
zero in the SM. These operators can be expressed in terms of
independent Lorentz structures. For example, there are four Lorentz
invariant structures for the lowest dimension $WW\gamma\gamma$ and
$ZZ\gamma\gamma$ interactions

\begin{eqnarray}
\textit{W}_{0}^{\gamma}=\frac{-e^{2}g^{2}}{2}F_{\mu \nu}F^{\mu \nu}
W^{+ \alpha} W_{\alpha}^{-},
\end{eqnarray}

\begin{eqnarray}
\textit{W}_{c}^{\gamma}=\frac{-e^{2}g^{2}}{4}F_{\mu \nu}F^{\mu
\alpha} (W^{+ \nu} W_{\alpha}^{-}+W^{- \nu} W_{\alpha}^{+}),
\end{eqnarray}

\begin{eqnarray}
\textit{Z}_{0}^{\gamma}=\frac{-e^{2}g^{2}}{4
\textmd{cos}^{2}\,\theta_{W}}F_{\mu \nu}F^{\mu \nu} Z^{\alpha}
Z_{\alpha},
\end{eqnarray}

\begin{eqnarray}
\textit{Z}_{c}^{\gamma}=\frac{-e^{2}g^{2}}{4
\textmd{cos}^{2}\,\theta_{W}}F_{\mu \nu}F^{\mu \alpha} Z^{\nu}
Z_{\alpha}.
\end{eqnarray}

Also, the two independent operators for the $ZZ\gamma\gamma$
interactions are parameterized as the following

\begin{eqnarray}
\textit{Z}_{0}^{Z}=\frac{-e^{2}g^{2}}{2
\textmd{cos}^{2}\,\theta_{W}}F_{\mu \nu}Z^{\mu \nu} Z^{\alpha}
Z_{\alpha},
\end{eqnarray}

\begin{eqnarray}
\textit{Z}_{c}^{Z}=\frac{-e^{2}g^{2}}{2
\textmd{cos}^{2}\,\theta_{W}}F_{\mu \nu}Z^{\mu \alpha} Z^{\nu}
Z_{\alpha}.
\end{eqnarray}

The five Lorentz structure belonging to $WWZ\gamma$ interactions are
given by

\begin{eqnarray}
\textit{W}_{0}^{Z}=-e^{2}g^{2} F_{\mu \nu}Z^{\mu \nu} W^{+ \alpha}
W_{\alpha}^{-},
\end{eqnarray}

\begin{eqnarray}
\textit{W}_{c}^{Z}=-\frac{e^{2}g^{2}}{2}F_{\mu \nu}Z^{\mu \alpha}
(W^{+ \nu} W_{\alpha}^{-}+W^{- \nu} W_{\alpha}^{+}),
\end{eqnarray}

\begin{eqnarray}
\textit{W}_{1}^{Z}=-\frac{e g_{z}g^{2}}{2}F^{\mu \nu} (W_{\mu
\nu}^{+}W_{\alpha}^{-} Z^{\alpha}+W_{\mu \nu}^{-}W_{\alpha}^{+}
Z^{\alpha}),
\end{eqnarray}

\begin{eqnarray}
\textit{W}_{2}^{Z}=-\frac{e g_{z}g^{2}}{2}F^{\mu \nu} (W_{\mu
\alpha}^{+} W^{-\alpha} Z_{\nu}+W_{\mu \alpha}^{-}W^{+ \alpha}
Z_{\nu}),
\end{eqnarray}

\begin{eqnarray}
\textit{W}_{3}^{Z}=-\frac{e g_{z}g^{2}}{2}F^{\mu \nu} (W_{\mu
\alpha}^{+} W^{-}_{\nu} Z^{\alpha}+W_{\mu \alpha}^{-}W_{\nu}^{+}
Z^{\alpha})
\end{eqnarray}
with $g=e/ \textmd{sin}\,\theta_{W}$, $g_{z}=e/
\textmd{sin}\,\theta_{W} \textmd{cos}\,\theta_{W}$ and
$V_{\mu\nu}=\partial_{\mu}V_{\nu}-\partial_{\nu}V_{\mu}$ where
$V=W^{\pm},Z$. Here, the CP-conserving anomalous $WW Z\gamma$ vertex
functions generated from Eqs. ($8$)-($12$) are given in
Appendix.

Consequently, these fourteen effective photonic quartic operators
can be simply expressed by

\begin{eqnarray}
\textit{L}=&&\frac{k_{0}^{\gamma}}{\Lambda^2}(\textit{Z}_{0}^{\gamma}+\textit{W}_{0}^{\gamma})+\frac{k_{c}^{\gamma}}{\Lambda^2}(\textit{Z}_{c}^{\gamma}
+\textit{W}_{c}^{\gamma})+\frac{k_{1}^{\gamma}}{\Lambda^2}\textit{Z}_{0}^{\gamma} \nonumber \\
&&+\frac{k_{23}^{\gamma}}{\Lambda^2}\textit{Z}_{c}^{\gamma}+\frac{k_{0}^{Z}}{\Lambda^2}\textit{Z}_{0}^{Z}+\frac{k_{c}^{Z}}{\Lambda^2}\textit{Z}_{c}^{Z}+\sum_{i=0,c,1,2,3}\frac{k_{i}^{W}}{\Lambda^2}\textit{W}_{i}^{Z}, \nonumber \\
\end{eqnarray}

where

\begin{eqnarray}
k_{j}^{\gamma}=k_{j}^{w}+k_{j}^{b}+k_{j}^{m}\,\,\,\,\,\,\,\,\,\,\,\,\,\,\,\,(j=0,c,1)
\end{eqnarray}

\begin{eqnarray}
k_{23}^{\gamma}=k_{2}^{w}+k_{2}^{b}+k_{2}^{m}+k_{3}^{w}+k_{3}^{m}
\end{eqnarray}

\begin{eqnarray}
k_{0}^{Z}=\frac{\textmd{cos}\,\theta_{W}}{\textmd{sin}\,\theta_{W}}(k_{0}^{w}+k_{1}^{w})-\frac{\textmd{sin}\,\theta_{W}}{\textmd{cos}\,\theta_{W}}(k_{0}^{b}+k_{1}^{b})+(\frac{\textmd{cos}^{2}\,\theta_{W}-\textmd{sin}^{2}\,\theta_{W}}{2\textmd{cos}\,\theta_{W}\textmd{sin}\,\theta_{W}})(k_{0}^{m}+k_{1}^{m}),
\end{eqnarray}

\begin{eqnarray}
k_{c}^{Z}=\frac{\textmd{cos}\,\theta_{W}}{\textmd{sin}\,\theta_{W}}(k_{c}^{w}+k_{2}^{w}+k_{3}^{w})-\frac{\textmd{sin}\,\theta_{W}}{\textmd{cos}\,\theta_{W}}(k_{c}^{b}+k_{2}^{b})+(\frac{\textmd{cos}^{2}\,\theta_{W}-\textmd{sin}^{2}\,\theta_{W}}{2\textmd{cos}\,\theta_{W}\textmd{sin}\,\theta_{W}})(k_{c}^{m}+k_{2}^{m}+k_{3}^{m}),
\end{eqnarray}

\begin{eqnarray}
k_{0}^{W}=\frac{\textmd{cos}\,\theta_{W}}{\textmd{sin}\,\theta_{W}}k_{0}^{w}-\frac{\textmd{sin}\,\theta_{W}}{\textmd{cos}\,\theta_{W}}k_{0}^{b}+(\frac{\textmd{cos}^{2}\,\theta_{W}-\textmd{sin}^{2}\,\theta_{W}}{2\textmd{cos}\,\theta_{W}\textmd{sin}\,\theta_{W}})k_{0}^{m},
\end{eqnarray}

\begin{eqnarray}
k_{c}^{W}=\frac{\textmd{cos}\,\theta_{W}}{\textmd{sin}\,\theta_{W}}k_{c}^{w}-\frac{\textmd{sin}\,\theta_{W}}{\textmd{cos}\,\theta_{W}}k_{c}^{b}+(\frac{\textmd{cos}^{2}\,\theta_{W}-\textmd{sin}^{2}\,\theta_{W}}{2\textmd{cos}\,\theta_{W}\textmd{sin}\,\theta_{W}})k_{c}^{m},
\end{eqnarray}

\begin{eqnarray}
k_{j}^{W}=k_{j}^{w}+\frac{1}{2}k_{j}^{m}\,\,\,\,\,\,\,\,\,\,\,\,\,\,\,\,(j=1,2,3).
\end{eqnarray}

For this study, we only take care of the $k_{i}^{W}$ parameters (see Eqs. ($18$)-($20$)) corresponding to the anomalous $WWZ\gamma$
couplings. These $k_{i}^{W}$ parameters are correlated with couplings defining anomalous $WW
\gamma \gamma, ZZ\gamma \gamma$ and $ZZZ \gamma$ interactions \cite{lhc}. Hence, we require to distinguish the anomalous $WWZ\gamma$
couplings from the other anomalous quartic couplings. This can be accomplished to apply extra
restrictions on $k_{i}^{j}$ parameters as suggested by Ref. \cite{lag3}. The anomalous $WWZ\gamma$ couplings can be only leaved by taking $k_{2}^{m}=-k_{3}^{m}$ while the remaining parameters are equal to zero. As a result, we express the effective interaction of $WWZ\gamma$
as follows
\begin{eqnarray}
\textit{L}_{eff}=\frac{k_{2}^{m}}{2\Lambda^{2}}(W_{2}^{Z}-W_{3}^{Z}).
\end{eqnarray}

Refs. \cite{lhc,lag3,lhc1} are phenomenologically investigated the $\frac{k_{2}^{m}}{\Lambda^{2}}$ couplings
defined the anomalous quartic $WWZ\gamma$ vertex. In addition, the $\frac{k_{0}^{W}}{\Lambda^{2}}$ and $\frac{k_{c}^{W}}{\Lambda^{2}}$ couplings
given in Eqs. ($18$)-($19$) constitute the
present experimental limits on the anomalous quartic $WWZ\gamma$ couplings within CP-conserving effective Lagrangians. Therefore, in this study, we examine limits on
the CP-conserving parameters $\frac{k_{0}^{W}}{\Lambda^{2}}$,
$\frac{k_{c}^{W}}{\Lambda^{2}}$, $\frac{k_{2}^{m}}{\Lambda^{2}}$ and the CP-violating parameter
$\frac{a_{n}}{\Lambda^{2}}$ to compare with the previous experimental and phenomenological results on the anomalous quartic $WWZ\gamma$ gauge couplings in the literature.

The anomalous quartic $WWZ\gamma$ couplings have been constrained by
analyzing the processes $e^{+}e^{-}\rightarrow
W^{+}W^{-}Z,W^{+}W^{-}\gamma,W^{+}W^{-}(\gamma)\rightarrow 4f\gamma$
\cite{lin,linb,linc,lind,line}, $e \gamma\rightarrow W^{+}W^{-}e,
\nu_{e}W^{-}Z$ \cite{lag1,lin2} and $\gamma\gamma\rightarrow
W^{+}W^{-}Z$ \cite{lin3,lin4} at linear $e^{+}e^{-}$ colliders and
its operating modes of $e \gamma$ and $\gamma\gamma$. In addition,
the potential of the process $e^{+}e^{-} \rightarrow e^{+}\gamma^{*}
e^{-} \rightarrow e^{+} W^{-} Z \nu_{e}$ \cite{mur} by making use of
Equivalent Photon Approximation (EPA) at the CLIC to probe the
anomalous quartic $WWZ\gamma$ gauge couplings is examined. Finally,
a detailed analysis of anomalous $WWZ\gamma$ couplings at the LHC
have been analyzed through the processes $pp\rightarrow$
$W(\rightarrow j j) \gamma Z(\rightarrow\ell^{+}  \ell^{-})$
\cite{lhc} and $W(\rightarrow\ell \nu_{\ell}) \gamma
Z(\rightarrow\ell^{+} \ell^{-})$ \cite{lhc1}.  Up to now, in these
studies, even though the anomalous quartic $WWZ\gamma$ couplings are
investigated via either CP-violating or CP-conserving effective
Lagrangians, they are examined by using both effective Lagrangians
solely by Refs. \cite{lhc1,mur}.

The LEP provides current experimental limits on $a_{n}/\Lambda^{2}$
parameter of the anomalous quartic $WWZ\gamma$ couplings determined
by CP-violating effective Lagrangian. Recent limits obtained through
the process $e^{+}e^{-}\rightarrow W^{+}W^{-} \gamma$ by L3, OPAL
and DELPHI collaborations are

\begin{eqnarray}
-0.14\,  \textmd{GeV}^{-2}<\frac{a_{n}}{\Lambda^{2}}<0.13\,\textmd{GeV}^{-2},
\end{eqnarray}

\begin{eqnarray}
-0.16\,  \textmd{GeV}^{-2}<\frac{a_{n}}{\Lambda^{2}}<0.15\,  \textmd{GeV}^{-2},
\end{eqnarray}

\begin{eqnarray}
-0.18\,  \textmd{GeV}^{-2}<\frac{a_{n}}{\Lambda^{2}}<0.14\,  \textmd{GeV}^{-2}
\end{eqnarray}
at $95\%$ confidence level, respectively \cite{lep1,lep2,lep3}.
Nevertheless, the most stringent limits on $k_{0}^{W}/\Lambda^{2}$
and $k_{c}^{W}/\Lambda^{2}$ parameters described by CP-conserving
effective Lagrangian are provided through the process
$q\overline{q}'\rightarrow W (\rightarrow\ell\nu)Z(\rightarrow jj)
\gamma$ with an integrated luminosity of $19.3$ fb$^{-1}$ at
$\sqrt{s}=8$ TeV by CMS collaboration at the LHC \cite{sınır}.
These are

\begin{eqnarray}
-1.2\times 10^{-5}  \textmd{GeV}^{-2}<\frac{k_{0}^{W}}{\Lambda^{2}}<1\times 10^{-5} \textmd{GeV}^{-2}
\end{eqnarray}
and
\begin{eqnarray}
-1.8\times 10^{-5}  \textmd{GeV}^{-2}<\frac{k_{c}^{W}}{\Lambda^{2}}<1.7\times 10^{-5} \textmd{GeV}^{-2}.
\end{eqnarray}
In the coming years, since the LHC will be upgraded to
center-of-mass energy of $14$ TeV, it is anticipated to introduce
more restrictive limits on anomalous quartic gauge boson couplings.

Photon-induced processes were comprehensively examined in $e p$ and
$e^{+} e^{-}$ collisions at the HERA and LEP, respectively. In
addition to $pp$ collisions at the LHC, photon-induced processes,
namely $\gamma \gamma$ and $\gamma p$, enable us to test of the
physics within and beyond the SM. These processes occurring at
center-of-mass energies well beyond the electroweak scale are
examined in an exactly undiscovered regime of the LHC. Although $p
p$ processes at the LHC reach very high effective luminosity, they
do not a clean environment due to the remnants of both proton beams
after the collision. On the other hand, since $\gamma \gamma$ and
$\gamma p$ processes have better known initial conditions and much
simpler final states, these interactions can compensate the
advantages of $p p$ processes. Initial state photons in $\gamma
\gamma$ and $\gamma p$ processes can be described in the framework
of the EPA \cite{esdeger}. In the EPA, while $\gamma \gamma$
collisions are generated by two almost real photons emitted from
protons, $\gamma p$ collisions are produced by one almost real
photon emitted from one incoming proton which then subsequently
collides with the other proton. The emitted photons in these
collisions have a low virtuality. Therefore, when a proton emits an
almost real photon, it does not dissociate into partons. Almost real
photons are scattered at very small angles from the beam pipe, and
they carry a small transverse momentum. Furthermore, if the proton
emits a photon, it scatters with a large pseudorapidity and can not
be detected from the central detectors. Hence, detection of intact
protons requires forward detector equipment in addition to central
detectors with large pseudorapidity providing some information on
the scattered proton energy. For this purpose, ATLAS and CMS
collaborations have a program of forward physics with extra
detectors located at $220$ m and $420$ m away from the interaction
point which can detect the particles with large pseudorapidity
\cite{il1,il2}. Forward detectors can detect intact scattered
protons with $9.5<\eta<13$ in a continuous range of $\xi$ where
$\xi$ is the proton momentum fraction loss described by
$\xi=(|\vec{p}|-|\vec{p}'|)/|\vec{p}|$; $\vec{p}$ and $\vec{p}'$ are
the momentum of incoming proton and the momentum of intact proton,
respectively. The relation between the transverse momentum and
pseudorapidity of intact proton is as follows

\begin{eqnarray}
p_{T}=\frac{\sqrt{E_{p}^{2}(1-\xi)^{2}-m_{p}^{2}}}{\cosh \eta}
\end{eqnarray}
where $m_{p}$ is the mass of proton and $E_{p}$ is the energy of
proton.

$\gamma \gamma$ collisions are usually electromagnetic in nature and
these reactions have less backgrounds compared to $\gamma p$
collisions. On the other hand, $\gamma p$ collisions can reach much
higher energy and effective luminosity with respect to $\gamma
\gamma$ collisions \cite{can1,can2}. These properties of $\gamma p$
process might be significant in the investigation of new physics due
to the high energy dependence of the cross section containing
anomalous couplings. Most of the SM operators are of dimension four
since only operators with even dimension satisfy conservation of
lepton and baryon number. Therefore, the operators examining
anomalous gauge boson self-couplings have to be at least dimension
six operators. For example, anomalous $WWZ\gamma$ couplings are
defined by dimension six effective Lagrangians, and have very strong
energy dependences. Hence anomalous cross section including the
$WWZ\gamma$ vertex has a higher momentum dependence than the SM
cross section. Therefore, $\gamma p$ processes are anticipated to
have a high sensitivity to anomalous $WWZ\gamma$ couplings since it
has a higher energy reach with respect to $\gamma \gamma$ process.

Photon-induced reactions were observed experimentally through the
processes $p\bar{p}\rightarrow p\gamma \gamma \bar{p}\rightarrow p
e^{+} e^{-} \bar{p}$ \cite{1,11}, $p\bar{p}\rightarrow p\gamma
\gamma \bar{p}\rightarrow p \mu^{+} \mu^{-} \bar{p}$ \cite{2},
$p\bar{p}\rightarrow p\gamma \bar{p}\rightarrow p W W \bar{p}$
\cite{WW} and $p\bar{p}\rightarrow p\gamma \bar{p}\rightarrow p
J/\psi (\psi(2S)) \bar{p}$ \cite{3} by the CDF and D0 collaborations
at the Fermilab Tevatron. However, after these processes were
examined at the Tevatron, this phenomenon has led to the
investigation of potential of the LHC as a $\gamma \gamma$ and
$\gamma p$ colliders for new physics researches. Therefore,
photon-photon processes such as $pp\rightarrow p\gamma \gamma
p\rightarrow p e^{+} e^{-} p$, $pp\rightarrow p\gamma \gamma
p\rightarrow p \mu^{+} \mu^{-} p$, and $pp\rightarrow p\gamma \gamma
p\rightarrow p W^{+} W^{-} p$ have been analyzed from the early LHC
data at $\sqrt{s}=7$ TeV by the CMS collaboration
\cite{CMS,CMS1,CMS2}. In addition, many studies on new physics
beyond the SM through photon-induced reactions at the LHC in the
literature have been phenomenologically examined. These studies
contain: gauge boson self-interactions, excited neutrino,
extradimensions, unparticle physics, and so forth
\cite{L1,L2,L3,Alboteanu:2008my,L4,L5,L6,L7,L8,L9,L10,L11,L12,L13,L14,L141,L15,L16,L17,L18,L19,L20,Sahin:2014dua,ban1,ban2,Fichet:2013ola}.
In this work, we have examined the CP-conserving and CP-violating
anomalous quartic $W W Z \gamma$ couplings through the process
$pp\rightarrow p\gamma p\rightarrow p W Z q' X$ at the LHC.

\section{The CROSS SECTIONS AND NUMERICAL ANALYSIS}
\begin{figure}
\includegraphics[width=0.5\columnwidth] {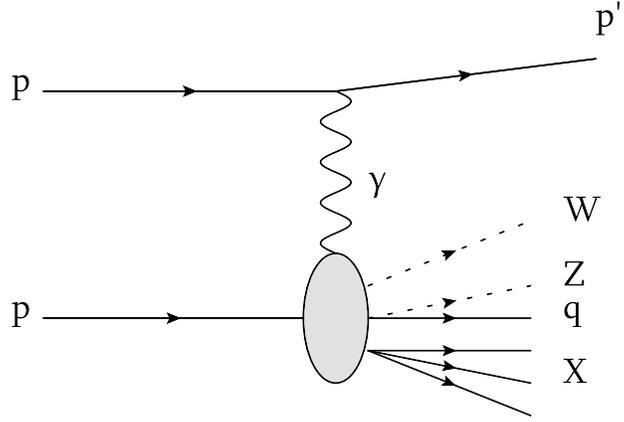}
\caption{Schematic diagram for the process $pp\rightarrow p\gamma
p\rightarrow p W Z q X$. \label{fig1}}
\end{figure}

\begin{figure}
\includegraphics[width=1\columnwidth] {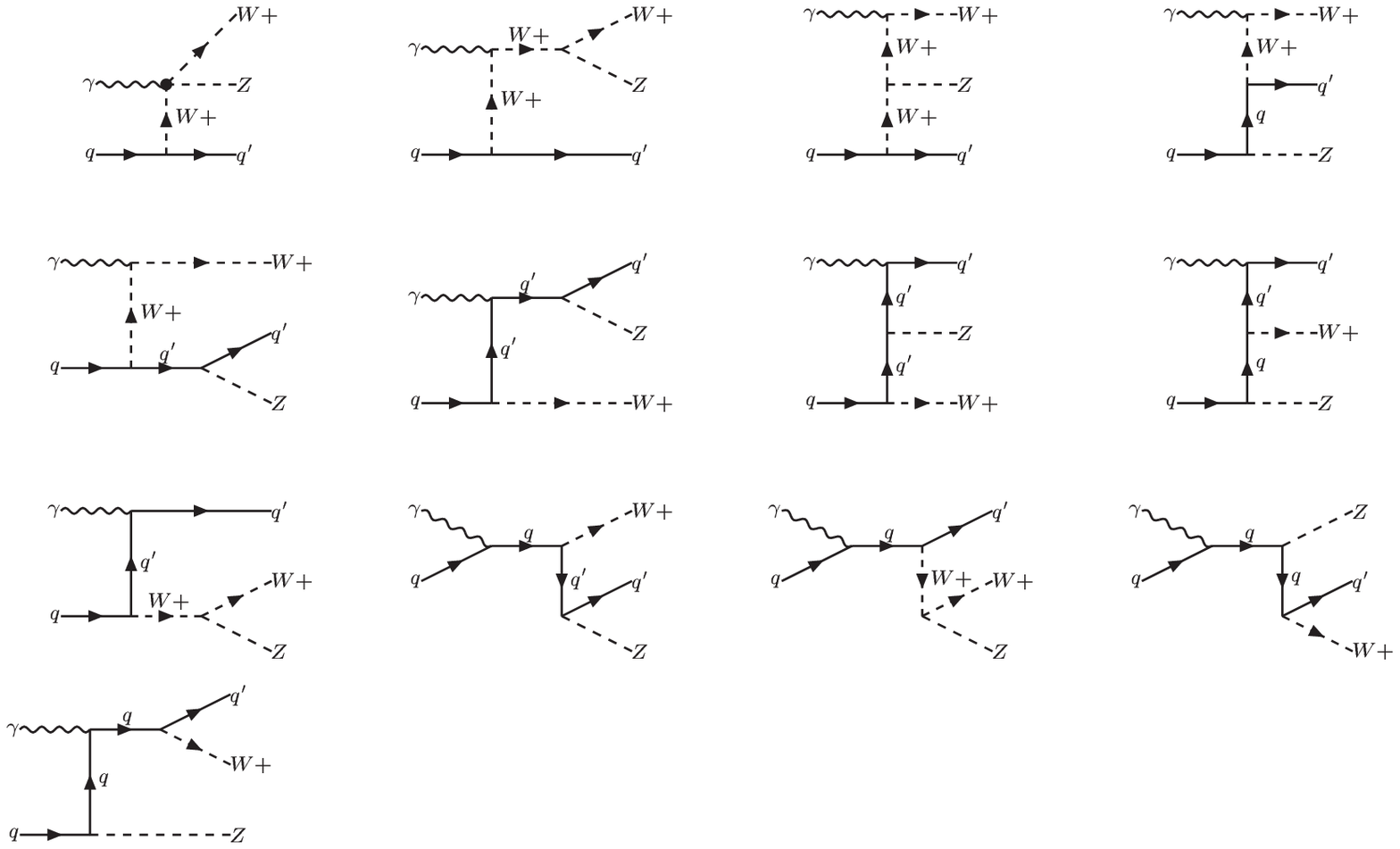}
\caption{Tree level Feynman diagrams for the subprocess $\gamma
q\rightarrow W Z q'$. \label{fig2}}
\end{figure}

An almost real photon emitted from one proton beam can interact with
the other proton and generate $W$ and $Z$ bosons via deep inelastic
scattering in the main process $pp\rightarrow p\gamma p\rightarrow p
W Z q' X$. A schematic diagram defining this main process is shown
in Fig. \ref{fig1}. The reaction $\gamma q\rightarrow W Z q'$
participates as a subprocess in the main process $pp\rightarrow
p\gamma p\rightarrow p W Z q' X$ where $q=d,s,\bar{u},\bar{c}$ and
$q'=u,c,\bar{d},\bar{s}$. Corresponding tree level Feynman diagrams
of the subprocess are shown in Fig. \ref{fig2}. As seen in Fig.
\ref{fig2}, while only the first of these diagrams includes
anomalous $WWZ\gamma$ vertex, the others give SM contributions. We
obtain the total cross section of $pp\rightarrow p\gamma
p\rightarrow p W Z q' X$ process by integrating differential cross
section of $\gamma q\rightarrow W Z q'$ subprocess over the parton
distribution functions CTEQ6L \cite{cte} and photon spectrum in EPA
by using the computer package CalcHEP \cite{calc}.

\begin{figure}
\includegraphics[width=0.8\columnwidth] {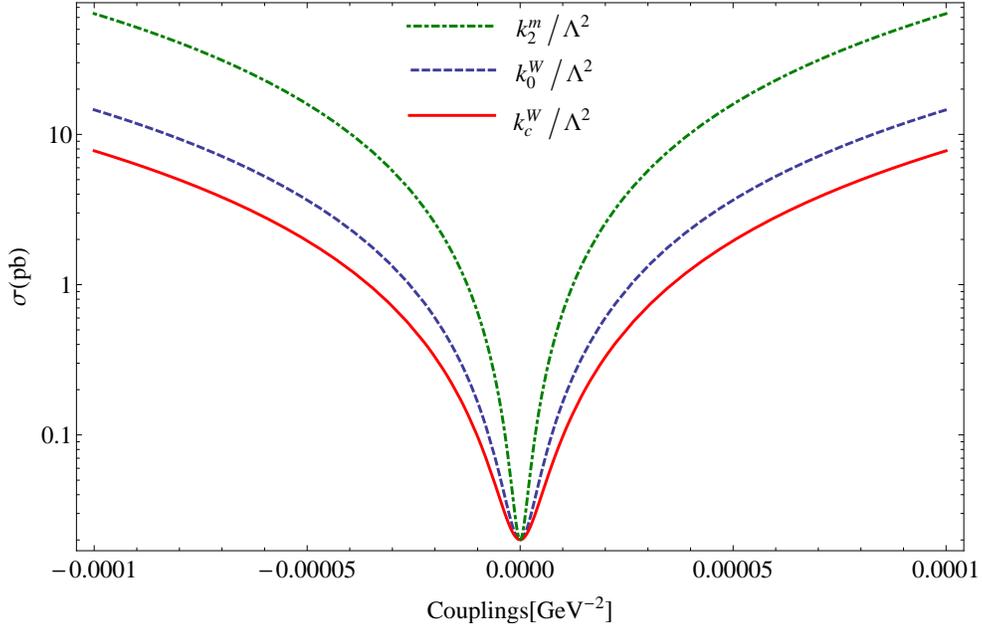}
\caption{The total cross sections as function of anomalous $\frac{k_{0}^{W}}{\Lambda^{2}}$, $\frac{k_{c}^{W}}{\Lambda^{2}}$ and $\frac{k_{2}^{m}}{\Lambda^{2}}$
couplings for the process $pp\rightarrow p\gamma p\rightarrow p W Z q X$ at the LHC with $\sqrt{s}=14$ TeV.
\label{fig3}}
\end{figure}

\begin{figure}
\includegraphics[width=0.8\columnwidth] {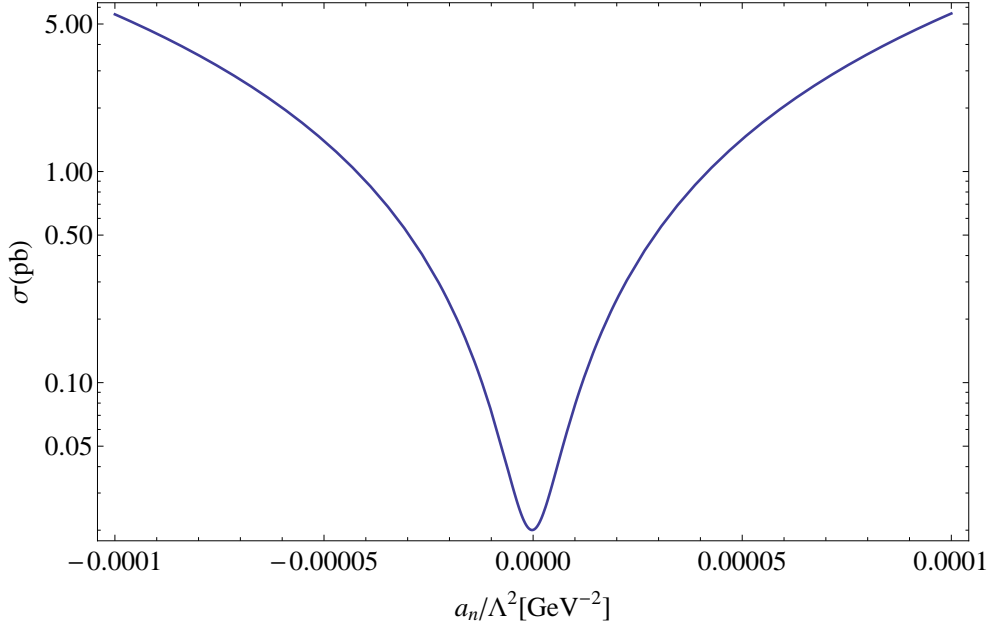}
\caption{The total cross section as function of anomalous
$\frac{a_{n}}{\Lambda^{2}}$ coupling for the process $pp\rightarrow
p\gamma p\rightarrow p W Z q X$ at the LHC with $\sqrt{s}=14$ TeV.
\label{fig4}}
\end{figure}

In Figs. \ref{fig3} and \ref{fig4}, we plot the integrated total
cross section of the process $pp\rightarrow p\gamma p\rightarrow p W
Z q' X$ as a function of the anomalous couplings. We collect all the
contributions arising from subprocesses $\gamma q\rightarrow W Z q'$
while obtaining the total cross section. In addition, we presume
that only one of the anomalous quartic gauge couplings is non-zero
at any given time, while the other couplings are fixed to zero. We
can see from Fig. \ref{fig3} that deviation from SM value of the
anomalous cross section containing the coupling
$\frac{k_{2}^{m}}{\Lambda^{2}}$ is larger than
$\frac{k_{0}^{W}}{\Lambda^{2}}$ and $\frac{k_{c}^{W}}{\Lambda^{2}}$.
For this reason, the limits obtained on the coupling
$\frac{k_{2}^{m}}{\Lambda^{2}}$ from analyzed process are
anticipated to be more restrictive than the limits on
$\frac{k_{0}^{W}}{\Lambda^{2}}$ and $\frac{k_{c}^{W}}{\Lambda^{2}}$.

We calculate the sensitivity of the process $pp\rightarrow p\gamma
p\rightarrow p W Z q' X$ to anomalous quartic gauge couplings by
applying one and two-dimensional $\chi^{2}$ criterion without a
systematic error. The $\chi^{2}$ function is defined as follows

\begin{eqnarray}
\chi^{2}=\left(\frac{\sigma_{SM}-\sigma_{NP}}{\sigma_{SM}\delta_{stat}}\right)^{2}
\end{eqnarray}
where $\sigma_{NP}$ is the cross section in the existence of new
physics effects, $\delta_{stat}=\frac{1}{\sqrt{N}}$ is the
statistical error: $N$ is the number of events. The number of
expected events of the process $pp\rightarrow p\gamma p\rightarrow p
W Z q' X$ is obtained as the signal $N=L_{int} \times \sigma_{SM}
\times S \times BR(W\rightarrow \ell \nu_{\ell})\times
BR(Z\rightarrow q\bar{q}')$ where $L_{int}$ denotes the integrated
luminosity, $\sigma_{SM}$ is the SM cross section and $\ell=e^{-}$
or $\mu^{-}$. We consider strong interactions between the
interacting protons. These interactions are generally performed by
adding a correction factor to the integrated cross section, which is
called the survival probability. Survival probability ($S$) is
described as the probability of the scattered protons not to
dissociate due to the secondary interactions. This survival
probability factor proposed for the some photoproduction processes
is $S=0.7$ \cite{L13,ban1,ban2}. The same survival factor is assumed
for our process. We impose both cuts for transverse momentum of
final state quarks to be $p_T^j >$ 15 GeV and the pseudorapidity of
final state quarks to be $|\eta|<2.5$ since ATLAS and CMS have
central detectors with a pseudorapidity coverage $|\eta|<2.5$. The
minimal transverse momentum cut of an outgoing proton is taken to be
$p_T >$ 0.1 GeV within the photon spectrum. After applying these
cuts, the SM background cross section for the process $pp\rightarrow
p\gamma p\rightarrow p W Z q X$ at $\sqrt{s}=14$ TeV is obtained as
0.0201 pb.

For the processes with the high luminosity at the LHC, physics
events called pile-up can give rise to an important background.
However, in low luminosity values the pile-up of events is
negligible in photoproduction interactions at the LHC. On the other
hand, the LHC using some of the techniques (kinematics and timing
constraints) can be operated at high luminosity such as $100$
fb$^{-1}$ as stated by Ref. \cite{il2}.

\begin{figure}
\includegraphics [width=0.8\columnwidth] {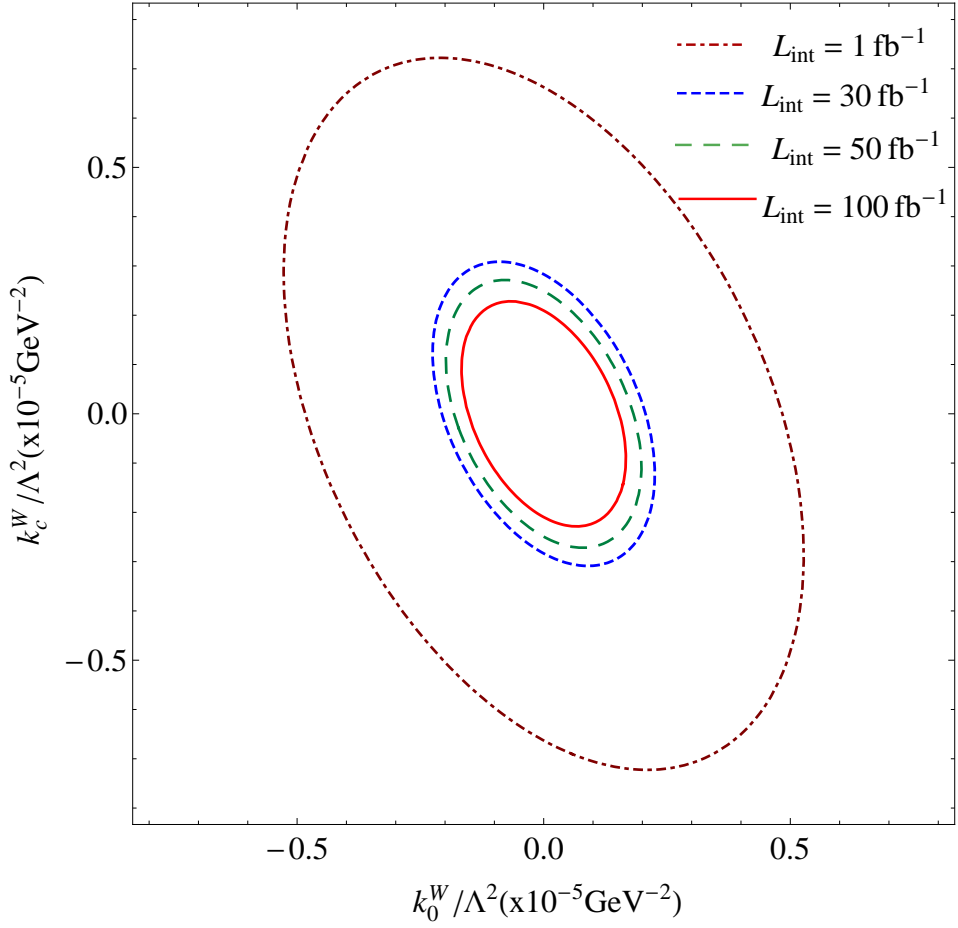}
\caption{$95\%$ C.L. contours for anomalous
$\frac{k_{0}^{W}}{\Lambda^{2}}$ and $\frac{k_{c}^{W}}{\Lambda^{2}}$
couplings for the process $pp\rightarrow p\gamma p\rightarrow p W Z
q X$ at the LHC with $\sqrt{s}=14$ TeV. \label{fig5}}
\end{figure}

In Tables \ref{tab1}-\ref{tab3}, we give the one-dimensional limits
on anomalous quartic gauge couplings
$\frac{k_{0}^{W}}{\Lambda^{2}}$, $\frac{k_{c}^{W}}{\Lambda^{2}}$,
$\frac{k_{2}^{m}}{\Lambda^{2}}$ and $\frac{a_{n}}{\Lambda^{2}}$ at
$95\%$ C.L. sensitivity at some integrated luminosities. Here, we
consider that only one of the anomalous couplings changes at any
time and center-of-mass energy of the $pp$ system is taken to be
$\sqrt{s}=14$ TeV. As can be seen from Tables, our limits obtained
on the couplings $\frac{k_{0}^{W}}{\Lambda^{2}}$,
$\frac{k_{c}^{W}}{\Lambda^{2}}$ and $\frac{a_{n}}{\Lambda^{2}}$ are
at the order of $10^{-6}$ GeV$^{-2}$ while limits on
$\frac{k_{2}^{m}}{\Lambda^{2}}$ are at the order of $10^{-7}$
GeV$^{-2}$. In addition, it can be understood that limits on the
coupling $\frac{k_{2}^{m}}{\Lambda^{2}}$ are more restrictive than
the limits on the couplings $\frac{k_{0}^{W}}{\Lambda^{2}}$ and
$\frac{k_{c}^{W}}{\Lambda^{2}}$. The sensitivities of the anomalous
couplings in $\frac{k_{0}^{W}}{\Lambda^{2}}$ -
$\frac{k_{c}^{W}}{\Lambda^{2}}$ plane at $\sqrt{s}=14$ TeV for
various integrated luminosities are shown in Fig. \ref{fig5}. As we
can see from Fig. \ref{fig5}, the best limits on anomalous couplings
$\frac{k_{0}^{W}}{\Lambda^{2}}$ and $\frac{k_{c}^{W}}{\Lambda^{2}}$
at $L_{int}=100$ fb$^{-1}$ and $\sqrt{s}=14$ TeV are obtained as
$[-1.66; 1.66]\times 10^{-6}$ GeV$^{-2}$ and $[-2.88; 1.88]\times
10^{-6}$ GeV$^{-2}$, respectively.

The topology of photon-induced interactions can separately take
place in diffractive processes. Diffractive processes are
characterized by the exchange of a colorless composite object called
as the pomeron. One of interactions including pomeron exchanges is
single diffraction interaction. Therefore, we can consider single
diffraction processes as a background of the analyzed process. A
pomeron emitted by any of the incoming proton immediately after it
collides with the other proton's quarks and this can produce same
final state particles. In deep inelastic scattering process the
virtuality of the struck quark is quite high. In our study, we take
the virtuality of the struck quark $Q^{2}=m_{Z}^{2}$ where $m_{Z}$
represents the $Z$ boson's mass. For this reason, when a pomeron
collides with a quark it may be dissociate into partons. These
interactions generally culminate in higher multiplicities of final
state particle due to existence of pomeron remnants \cite{can1}.
Hence, pomeron remnants can be determined by the calorimeters and
this background can be removed. In addition, the survival
probability for a pomeron exchange is quite smaller than the
survival probability of induced photons \cite{can2}. Hence, even
though the background arising from pomeron-induced process are not
annihilated, it can not be too large with respect to the SM
background contributions coming than the photon-induced process. It
can be supposed that even if the background contribution of
pomeron-induced process to our analyzed process is up to the SM
background, all our limits with a $100$ fb$^{-1}$ of integrated
luminosity at $\sqrt{s}=14$ TeV are broken up to $3$ times.
\section{Conclusions}

The LHC with forward detector equipment is a suitable platform to
examine physics within and beyond the SM via $\gamma \gamma$ and
$\gamma p$ processes. $\gamma p$ process has the high luminosities
and high center-of-mass energies compared to $\gamma \gamma$
process. Moreover, $\gamma p$ process due to the remnants of only
one of the proton beams provides rather clean experimental
conditions according to pure deep inelastic scattering of $pp$
process. For these reasons, we examine the process $pp\rightarrow
p\gamma p\rightarrow p W Z q X$ in order to determine anomalous
quartic $WWZ\gamma$ parameters $\frac{k_{0}^{W}}{\Lambda^{2}}$,
$\frac{k_{c}^{W}}{\Lambda^{2}}$, $\frac{k_{2}^{m}}{\Lambda^{2}}$ and
$\frac{a_{n}}{\Lambda^{2}}$ obtained by using two different
CP-violating and CP-conserving effective Lagrangians at the LHC. A
featured advantage of the process $pp\rightarrow p\gamma
p\rightarrow p W Z q X$ is that it isolates anomalous $WWZ \gamma$
couplings. It enables us to probe $WWZ \gamma$ couplings independent
of $WW \gamma \gamma$. Our limits on $\frac{k_{0}^{W}}{\Lambda^{2}}$
and $\frac{k_{c}^{W}}{\Lambda^{2}}$ are approximately one order
better than the LHC's limits \cite{sınır} while the limits
obtained on $\frac{a_{n}}{\Lambda^{2}}$ can set more stringent limit
by five orders of magnitude compared to LEP results \cite{lep1}.
Moreover, we compare our limits with phenomenological studies on the
anomalous $\frac{k_{2}^{m}}{\Lambda^{2}}$ and
$\frac{a_{n}}{\Lambda^{2}}$ couplings at the LHC and CLIC. Ref.
\cite{mur} have considered semi-leptonic decay channel of the final
$W$ and $Z$ bosons in the cross section calculations to improve the
limits on anomalous $\frac{a_{n}}{\Lambda^{2}}$ and
$\frac{k_{2}^{m}}{\Lambda^{2}}$ couplings at the CLIC. We can see
that the limits on anomalous $\frac{a_{n}}{\Lambda^{2}}$ and
$\frac{k_{2}^{m}}{\Lambda^{2}}$ couplings expected to be obtained
with $L_{int}=590$ fb$^{-1}$ and $\sqrt{s}=3$ TeV are almost $3$
times better than our best limits. Nevertheless, the limits on
$\frac{a_{n}}{\Lambda^{2}}$ by Ref. \cite{lhc1} have derived through
$W$ and $Z$'s pure leptonic decays at the LHC $14$ TeV with $100$
fb$^{-1}$. Our best limit is $10$ times more restrictive than the
best limit obtained in Ref. \cite{lhc1}.

\pagebreak

\appendix*
\section{The anomalous vertex functions for $W W Z \gamma$}

The anomalous vertex for $W^{+} (p_{1}^{\alpha}) W^{-}(p_{2}^{\beta}) Z (k_{2}^{\nu}) \gamma (k_{1}^{\mu})$ with the help of effective
Lagrangian Eq. ($1$) is generated as follows

\begin{eqnarray}
&&i\frac{\pi\alpha}{4 \textmd{cos}\,\, \theta_{W}\Lambda^{2}}a_{n}[g_{\alpha\nu}[g_{\beta\mu}\, k_{1}.(k_{2}-p_{1})-k_{1\beta}.(k_{2}-p_{1})_{\mu}] \nonumber \\
&&-g_{\beta\nu}[g_{\alpha\mu}\, k_{1}.(k_{2}-p_{2})-k_{1\alpha}.(k_{2}-p_{2})_{\mu}] \nonumber \\ &&+g_{\alpha\beta}[g_{\nu\mu}k_{1}.(p_{1}-p_{2})-k_{1\nu}.(p_{1}-p_{2})_{\mu}] \nonumber \\
&&-k_{2\alpha}(g_{\beta\mu}k_{1\nu}-g_{\nu\mu}k_{1\beta})+k_{2\beta}(g_{\alpha\mu}k_{1\nu}-g_{\nu\mu}k_{1\alpha}) \nonumber \\
&&-p_{2\nu}(g_{\alpha\mu}k_{1\beta}-g_{\beta\mu}k_{1\alpha})+p_{1\nu}(g_{\beta\mu}k_{1\alpha}-g_{\alpha\mu}k_{1\beta}) \nonumber \\
&&+p_{1\beta}(g_{\nu\mu}k_{1\alpha}-g_{\alpha\mu}k_{1\nu})+p_{2\alpha}(g_{\nu\mu}k_{1\beta}-g_{\beta\mu}k_{1\nu})]. \nonumber \\
\end{eqnarray}

In addition, the vertex functions for $W^{+} (p_{1}^{\alpha})
W^{-}(p_{2}^{\beta}) Z (k_{2}^{\nu}) \gamma (k_{1}^{\mu})$ produced
from the effective Lagrangians Eqs. ($8$)-($12$) are expressed below

\begin{eqnarray}
2ie^{2}g^{2}g_{\alpha\beta}[g_{\mu\nu}(k_{1}.k_{2})-k_{1\nu}k_{2\mu}],
\end{eqnarray}

\begin{eqnarray}
&&i\frac{e^{2}g^{2}}{2}[(g_{\mu\alpha}g_{\nu\beta}+g_{\nu\alpha}g_{\mu\beta})(k_{1}.k_{2})+g_{\mu\nu}(k_{2\beta}k_{1\alpha}+k_{1\beta}k_{2\alpha}) \nonumber \\
&&-k_{2\mu}k_{1\alpha}g_{\nu\beta}-k_{2\beta}k_{1\nu}g_{\mu\alpha}-k_{2\alpha}k_{1\nu}g_{\mu\beta}-k_{2\mu}k_{1\beta}g_{\nu\alpha}].
\end{eqnarray}

\begin{eqnarray}
ieg_{z}g^{2}((g_{\mu \alpha}k_{1}.p_{1}-p_{1\mu}k_{1\alpha})g_{\nu
\beta}+(g_{\mu \beta}k_{1}.p_{2} - p_{2\mu}k_{1\beta})g_{\nu
\alpha})
\end{eqnarray}

\begin{eqnarray}
&&i\frac{eg_{z}g^{2}}{2}((k_{1}.p_{1}+k_{1}.p_{2})g_{\mu \nu}g_{\alpha \beta}-(k_{1 \alpha}p_{1\beta}+k_{1\beta}p_{2\alpha})g_{\mu \nu} \nonumber \\
&&-(p_{1\mu}+p_{2\mu})k_{1\nu}g_{\alpha \beta}+(p_{1\beta}g_{\mu\alpha}+p_{2\alpha}g_{\mu \beta})k_{1\nu})
\end{eqnarray}

\begin{eqnarray}
&&i\frac{eg_{z}g^{2}}{2}(k_{1}.p_{1}g_{\mu \beta}g_{\nu \alpha}+k_{1}.p_{2}g_{\mu \alpha}g_{\nu \beta}+(p_{1\nu}-p_{2\nu})k_{1\beta} g_{\mu \alpha} \nonumber \\
&& -(p_{1\nu}-p_{2\nu})k_{1\alpha} g_{\mu \beta}-p_{1\mu}k_{1\beta}g_{\nu \alpha}-p_{2\mu}k_{1\alpha}g_{\nu \beta}).
\end{eqnarray}

\pagebreak

\pagebreak


\begin{table}
\caption{$95\%$ C.L. sensitivity limits of the anomalous
$\frac{k_{0}^{W}}{\Lambda^{2}}$ and $\frac{k_{c}^{W}}{\Lambda^{2}}$
couplings through the process $pp\rightarrow p\gamma p\rightarrow p
W Z q X$. Here, center-of-mass energy of the $pp$ system is taken to
be $\sqrt{s}=14$ TeV. \label{tab1}}
\begin{ruledtabular}
\begin{tabular}{ccc}
$L_{int}$(fb$^{-1}$)& $\frac{k_{0}^{W}}{\Lambda^{2}}$(GeV$^{-2}$)& $\frac{k_{c}^{W}}{\Lambda^{2}}$ (GeV$^{-2}$))\\
\hline
$1$& $[-4.33; 4.33]\times 10^{-6}$& $[-5.93; 5.93]\times 10^{-6}$ \\

$30$& $[-1.85; 1.85]\times 10^{-6}$& $[-2.53; 2.53]\times 10^{-6}$ \\

$50$&  $[-1.63; 1.63]\times 10^{-6}$& $[-2.23; 2.23]\times 10^{-6}$ \\

$100$&  $[-1.37; 1.37]\times 10^{-6}$& $[-1.88; 1.88]\times 10^{-6}$ \\
\end{tabular}
\end{ruledtabular}
\end{table}

\begin{table}
\caption{$95\%$ C.L. sensitivity limits of the anomalous
$\frac{k_{2}^{m}}{\Lambda^{2}}$ couplings through the process
$pp\rightarrow p\gamma p\rightarrow p W Z q X$. Here, center-of-mass
energy of the $pp$ system is taken to be $\sqrt{s}=14$ TeV.
\label{tab2}}
\begin{ruledtabular}
\begin{tabular}{cc}
$L_{int}$(fb$^{-1}$)& $\frac{k_{2}^{m}}{\Lambda^{2}}$(GeV$^{-2}$)\\
\hline
$1$& $[-2.07; 2.07]\times 10^{-6}$ \\

$30$& $[-8.85; 8.85]\times 10^{-7}$  \\

$50$&  $[-7.79; 7.79]\times 10^{-7}$\\

$100$&  $[-6.55; 6.55]\times 10^{-7}$ \\
\end{tabular}
\end{ruledtabular}
\end{table}

\begin{table}
\caption{$95\%$ C.L. sensitivity limits of the anomalous
$\frac{a_{n}}{\Lambda^{2}}$ couplings through the process
$pp\rightarrow p\gamma p\rightarrow p W Z q X$. Here, center-of-mass
energy of the $pp$ system is taken to be $\sqrt{s}=14$ TeV.
\label{tab3}}
\begin{ruledtabular}
\begin{tabular}{cc}
$L_{int}$(fb$^{-1}$)& $\frac{a_{n}}{\Lambda^{2}}$ (GeV$^{-2}$)\\
\hline
$1$&  $[-7.00; 7.00]\times 10^{-6}$ \\

$30$&  $[-2.99; 2.99]\times 10^{-6}$  \\

$50$&  $[-2.63; 2.63]\times 10^{-6}$\\

$100$&  $[-2.21; 2.21]\times 10^{-6}$ \\
\end{tabular}
\end{ruledtabular}
\end{table}

\end{document}